# Audited calibration under regime shift as a computational test of support-structured broadcast

Mark Walsh, PharmD
Corresponding Email: mark.walsh@rutgers.edu

**Abstract**

A central prediction of the accompanying theoretical framework is that metacognitive calibration can vary even when content-level performance is held approximately fixed, depending on whether support structure is preserved in a globally reusable broadcast state. We provide a minimal computational test of this claim using a two-channel probabilistic cue-integration task with regime shifts that induce systematic miscalibration in one channel. We compare content-dominated architectures, in which confidence is calibrated by a single global mapping from evidence strength to probability, to an auditor architecture that learns a regime-conditioned calibration mapping from an audit trail of outcomes. We then couple confidence to control by implementing a policy that either acts immediately or requests one additional sample when confidence falls below a threshold. Across matched evidence streams, the auditor substantially improves calibration, particularly in the degraded regime, and produces qualitatively different control behavior by selectively requesting additional evidence under low-support conditions. These results demonstrate a concrete, testable dissociation between content performance and system-level confidence and policy that arises from globally reusable support summaries.

## 1. Introduction

The companion theoretical manuscript proposes that global broadcast can preserve compact support structure alongside content, yielding a presentation profile that can be reused for calibration and control (Walsh 2026). A key empirical prediction follows: when content discrimination is held approximately constant, manipulations that change the availability of support structure should alter subjective quality, metacognitive calibration, and downstream policy.

Here we test the computational core of that prediction in a minimal setting. We implement a probabilistic cue-integration model with two evidence channels and a latent regime variable that changes the reliability of one channel. The content decision rule is shared across model families, but their confidence and control policies differ depending on whether a global controller (an auditor) has access to a compact support summary identifying the regime. This directly targets the proposed distinction between content-dominated broadcast and content-plus-presentation broadcast.

Motivating claim (from the companion framework):

> *"The key comparison is between models in which evidence and vehicle variables remain local gain parameters and models in which compact presentation profiles are explicitly represented and accessible to a global control layer that performs auditor-like functions ... Once accuracy is matched, one can probe behavior under controlled perturbations of vehicle or evidence structure ... Models with auditor functions should also show system-level confidence that differs systematically from constituent module confidences, in a way that improves calibration relative to simple aggregation."*

We operationalize "audit trail" as outcome-conditioned learning of a confidence mapping, and we operationalize control consequences by allowing the system to request one additional sample when system-level confidence is low. This yields a clean computational instantiation of the claim that globally reusable support summaries can create variance in outbound policy, not merely in internal confidence.

Post hoc calibration and uncertainty estimation are well established in machine learning, with temperature scaling as a standard baseline for improving probabilistic forecasts without changing the underlying classifier (Guo et al. 2017). A large literature also emphasizes that predictive confidence can degrade under distribution shift and nonstationary noise, motivating methods and benchmarks for robustness and uncertainty under shift (Ovadia et al. 2019; Hendrycks and Gimpel 2018). In parallel, decision systems with abstention, selective prediction, and reject options formalize the idea that an agent can trade coverage for reliability, and these ideas extend naturally to sequential settings where an agent can gather additional evidence when uncertain (Chow 1970; Geifman and El-Yaniv 2017). Within cognitive science and decision theory, sequential sampling models and confidence based control have long been used to connect evidence accumulation to action and information seeking (Ratcliff and McKoon 2008; Pleskac and Busemeyer 2010).

The present work is not intended to introduce a new calibration algorithm or a new reject option policy. Instead, it isolates a specific architectural contrast that is central to the companion theoretical proposal. We consider a regime shift that induces systematic miscalibration in one channel while keeping the content decision computation fixed across model families. Under this condition, a confidence threshold control policy is only as good as the confidence it uses. A content dominated mapping can remain confidently wrong under shift and therefore fail to request additional evidence when it is most valuable. We show that conditioning confidence on a compact support summary, and learning that mapping from outcome feedback via an audit trail, yields sharply improved calibration in the degraded regime and produces targeted information seeking behavior under the act versus sample policy. This connects calibration under shift to a concrete control signature while holding content inference constant, which is not typically the focus of existing calibration or selective prediction treatments.

We now define the generative task, model families, and evaluation metrics used to test this architectural contrast.

## 2. Methods

### 2.1 Task and generative process

We consider a binary latent state $X \in \{0,1\}$. On each trial, two channels provide noisy evidence about $X$. Channel A is stable. Channel B undergoes regime shifts that change its observation noise. Let $F \in \{good, bad\}$ denote the regime. Evidence samples are generated as $y_A = X + \varepsilon_A$ and $y_B = X + \varepsilon_B$ with $\varepsilon_A \sim N(0, \sigma_A^2)$ and $\varepsilon_B \sim N(0, \sigma_B(F)^2)$. The bad regime occurs with probability $p_{bad}$. Unless otherwise specified, we use $\sigma_A = 0.7$, $\sigma_{B,good} = 0.7$, $\sigma_{B,bad} = 2.0$, and $p_{bad} = 0.3$. In the notation of the companion framework, $X$ is content, the evidence samples $(y_A, y_B)$ constitute $E$, and the regime variable $F \in \{good, bad\}$ is a maximally simplified vehicle variable standing in for richer support summaries.

### 2.2 Content decision rule

All model families share the same content decision rule based on a log-likelihood ratio under an assumed observation model. For each channel $i \in \{A, B\}$ we compute a local log-likelihood ratio $\ell_i(y_i) = \log p(y_i|X = 1) - \log p(y_i|X = 0)$. The integrated log-odds is $L = \ell_A + \ell_B$, and the content decision is $\hat{X} = 1$ if $L \geq 0$ else 0. To induce a controlled miscalibration, the local computation for channel B assumes the good-regime noise $\sigma_{B,good}$ even when the environment is in the bad regime. This yields a systematic confidence error under regime shift while preserving a stable decision computation across families.

### 2.3 Explicit equations and reproducibility details

We assume a Gaussian observation model for each channel $i \in \{A, B\}$: $y_i \mid X = x \sim \mathcal{N}(x, \sigma_i^2)$, where $x \in \{0,1\}$. The local log-likelihood ratio is $\ell_i(y_i) = \log p(y_i|X = 1) - \log p(y_i|X = 0)$, which under the Gaussian model simplifies to $\ell_i(y_i) = (y_i^2 - (y_i - 1)^2)/(2\sigma_i^2) = (2y_i - 1)/(2\sigma_i^2)$. We compute the integrated log-odds as $L = \ell_A(y_A; \sigma_A) + \ell_B(y_B; \sigma_{B,assumed})$, where $\sigma_{B,assumed} = \sigma_{B,good}$ even in the bad regime.

Confidence mapping uses $p = \sigma(\alpha L) = 1/(1 + \exp(-\alpha L))$. The uncalibrated baseline uses $\alpha = 1$. The global temperature baseline fits a single $\alpha$ by minimizing NLL on the training split via grid search over $\alpha \in [0.05, 5.0]$ with 300 equally spaced grid points.

The auditor fits $\alpha_{good}$ and $\alpha_{bad}$ separately on training data conditioned on the true regime $F$.

ECE is computed using $B = 20$ equal-width bins over [0,1]. Let $n_k$ be the number of samples in bin $k$, $conf(k)$ the mean predicted probability in bin $k$, and $acc(k)$ the empirical frequency of $X = 1$ in bin $k$. Then $ECE = \Sigma_k (n_k/n) \cdot |acc(k) - conf(k)|$. All results use a 50/50 train/test split with $N = 250{,}000$ trials and fixed random seeds (NumPy default_rng seed 0) for dataset generation and second-sample generation; the online $\alpha_{bad}$ learning illustration uses seed 1 for selecting bad-regime samples.

### 2.4 Confidence Mapping and audit-trail learning (online)

Given integrated log-odds $L$, a system-level probability estimate is computed as $p = \sigma(\alpha L)$, where $\sigma(\cdot)$ is the logistic function and $\alpha$ is a temperature-like calibration parameter. We compare three mappings:

(i) Uncalibrated content-dominated: $\alpha = 1$.
(ii) Globally calibrated content-dominated: a single $\alpha$ is fitted on training data by minimizing negative log-likelihood.
(iii) Regime-aware auditor: separate $\alpha_{good}$ and $\alpha_{bad}$ are fitted on training data conditioned on the regime $F$. This corresponds to an auditor that has access to a compact support summary sufficient to discriminate regimes.

To illustrate audit-trail accumulation over time, we also simulate an online learning variant in which $\alpha_{bad}$ is updated via stochastic gradient descent on bad-regime samples. This demonstrates how a controller can adapt its calibration based on outcome feedback, producing systematic changes in confidence and policy in the degraded regime.

### 2.5 Control policy: act or request one more sample

We couple confidence to control by implementing a two-stage policy. After computing $p$ and confidence $c = max(p, 1 - p)$, the system either acts immediately if $c \geq \tau$, or requests one additional sample if $c < \tau$. Because confidence is miscalibrated under regime shift in the content-dominated models, this threshold policy can fail to request additional evidence precisely when it is most needed. If it requests a sample, a second independent draw $(y_{A2}, y_{B2})$ is generated under the same regime, a second log-odds $L2$ is computed, and the system updates its evidence by $L \leftarrow L + L2$ before acting. Requesting an additional sample incurs a fixed cost $\kappa$. Utility is defined as $+1$ for a correct final decision, $-1$ for an incorrect decision, and $-\kappa$ if a second sample is requested. Unless otherwise specified, $\tau = 0.8$ and $\kappa = 0.05$.

### 2.6 Evaluation metrics

We report content accuracy, negative log-likelihood (NLL), Brier score, and expected calibration error (ECE). For control, we report mean utility and the request rate (fraction of trials on which an additional sample is requested). All results are computed on a held-out test split (50% train, 50% test) with 250,000 trials total. We distinguish pre-policy DecisionAccuracy (shared across models) from post-policy FinalAccuracy under the act-versus-sample control policy.

To interpret these metrics in practical terms, accuracy reflects how often the system's final choice matches the true latent state. The calibration measures evaluate whether the system's stated probabilities are trustworthy. NLL and Brier score reward probabilistic forecasts that match empirical outcomes, and they penalize overconfident errors more than cautious ones. ECE is the most intuitive. If ECE is near zero, then predictions assigned a given confidence level tend to be correct at roughly that frequency (for example, among trials where the model reports about 0.8 probability, it is correct about 80% of the time). For control, the request rate reports how often the system chooses to gather one additional sample before acting, and mean utility summarizes the resulting tradeoff between improved decisions and the sampling cost. We report both pre-policy DecisionAccuracy, which isolates the shared content decision rule, and post-policy FinalAccuracy, which reflects the combined effect of confidence calibration and the act-versus-sample control policy.

## 3. Results

Across the same evidence stream and shared content decision rule, the auditor architecture achieves markedly improved calibration relative to content-dominated mappings, with the largest gains concentrated in the degraded regime. These calibration gains translate into systematic policy differences. The auditor requests additional samples more strongly in the degraded regime, producing a distinct act-versus-sample profile.

*Table 1. Calibration and policy metrics (overall and bad-regime subset). DecAcc reports the shared pre-policy decision rule; FinalAcc reflects the act-versus-sample policy ($\tau = 0.8$, $\kappa = 0.05$).*

| Model | Subset | DecAcc | NLL | ECE | Brier | FinalAcc | Sample | Utility |
|---|---|---|---|---|---|---|---|---|
| Uncalibrated | Overall | 0.7941 | 0.5538 | 0.0638 | 0.1541 | 0.8504 | 0.2978 | 0.6860 |
| Global temp | Overall | 0.7941 | 0.4864 | 0.0544 | 0.1529 | 0.8649 | 0.5604 | 0.7019 |
| Auditor | Overall | 0.7941 | 0.4253 | 0.0024 | 0.1391 | 0.8596 | 0.4777 | 0.6952 |
| Uncalibrated | Bad only | 0.6810 | 1.0175 | 0.2099 | 0.2528 | 0.7164 | 0.2262 | 0.4216 |
| Global temp | Bad only | 0.6810 | 0.6858 | 0.1285 | 0.2219 | 0.7343 | 0.4240 | 0.4474 |
| Auditor | Bad only | 0.6810 | 0.5910 | 0.0077 | 0.2030 | 0.7504 | 0.8181 | 0.4599 |

Table 1 shows that DecisionAccuracy is identical across model families, both overall and within the degraded regime, confirming that differences arise from confidence mapping and control rather than from different content inference. Calibration improves most strongly under regime shift. In the bad-regime subset, the auditor reduces ECE by more than an order of magnitude relative to the best content-dominated baseline, indicating that its stated probabilities track empirical correctness most closely in the regime where miscalibration is otherwise largest. These calibration differences translate into control. In the bad regime, the auditor requests an additional sample far more often, increasing FinalAccuracy despite higher sampling costs. Mean utility therefore reflects an explicit tradeoff between correctness and evidence cost under the chosen parameters, which we examine directly by sweeping $\kappa$ below.

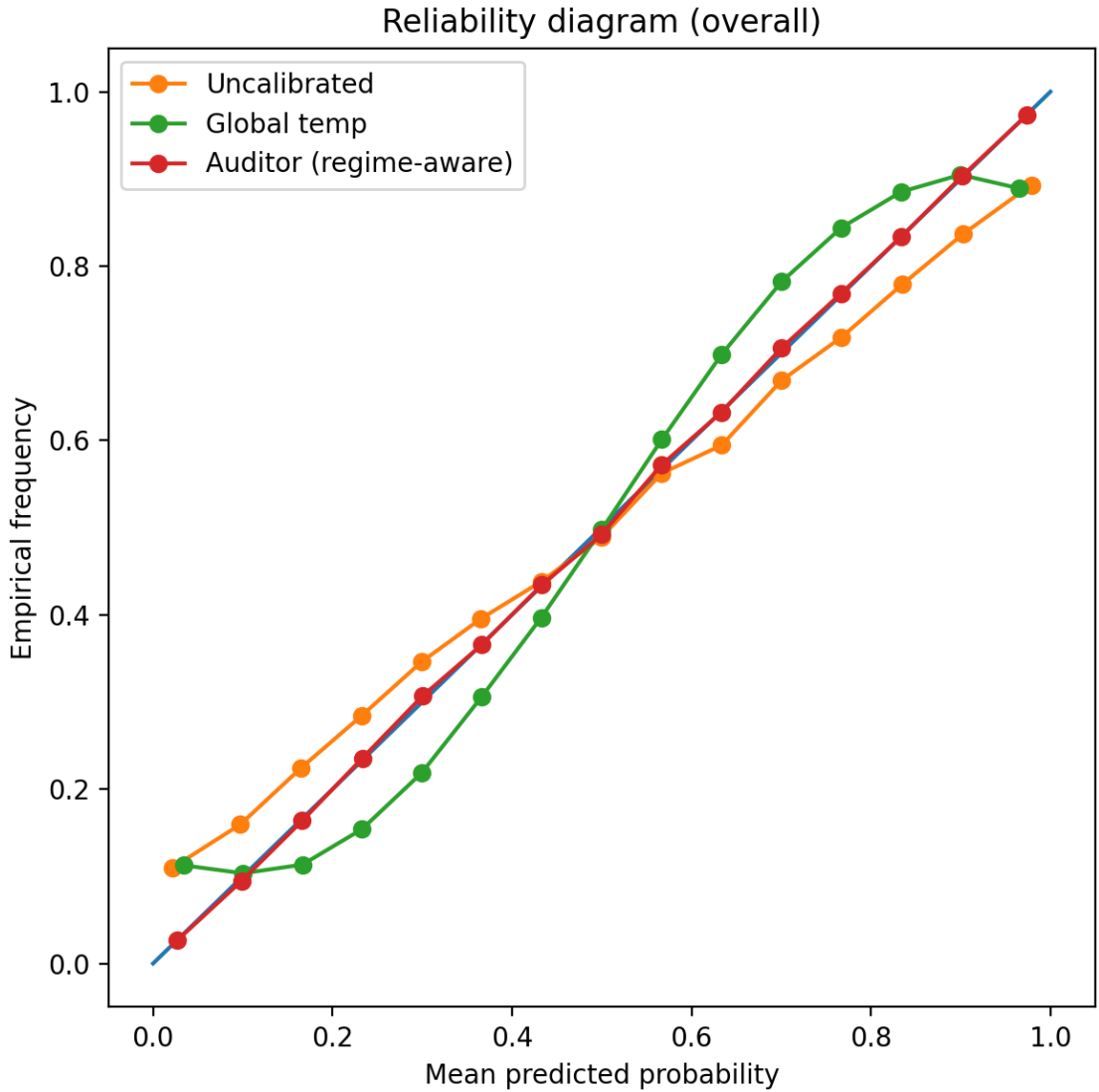

*Figure 1. Reliability diagram (overall).* *Predicted probability vs empirical correctness frequency, pooled across regimes.*

Figure 1 provides an overall calibration view. The auditor remains closest to the diagonal across bins, while the content-dominated mappings deviate systematically, consistent with the ECE differences in Table 1.

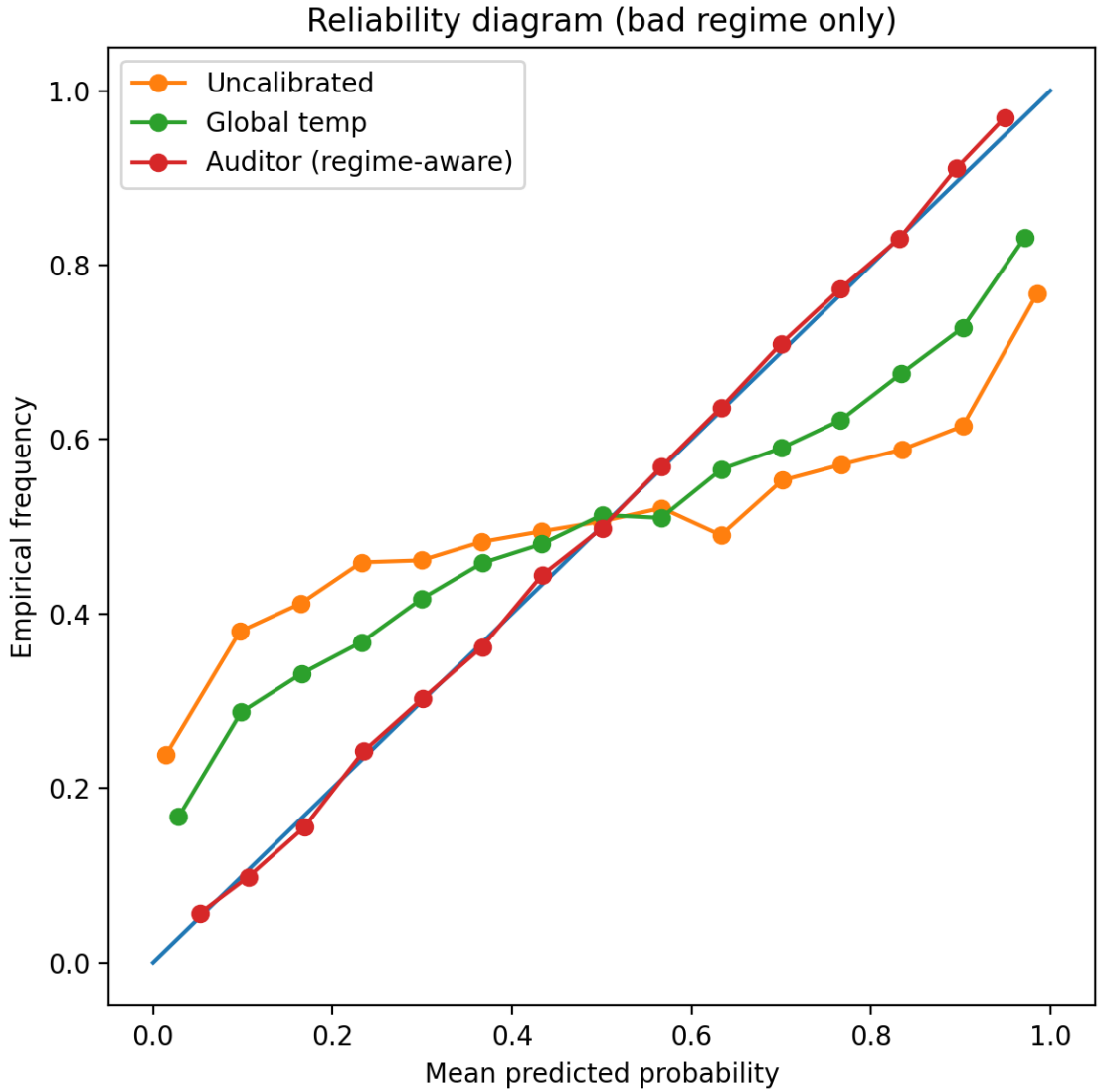

*Figure 2. Reliability diagram (bad regime only). Same as Figure 1, restricted to bad-regime trials.*

Figure 2 isolates the degraded regime, where miscalibration is most pronounced. Both content-dominated mappings exhibit clear overconfidence at higher predicted probabilities. Conditioning confidence on regime allows the auditor to remain near-diagonal across bins, demonstrating that its advantage is concentrated precisely where reliability shifts.

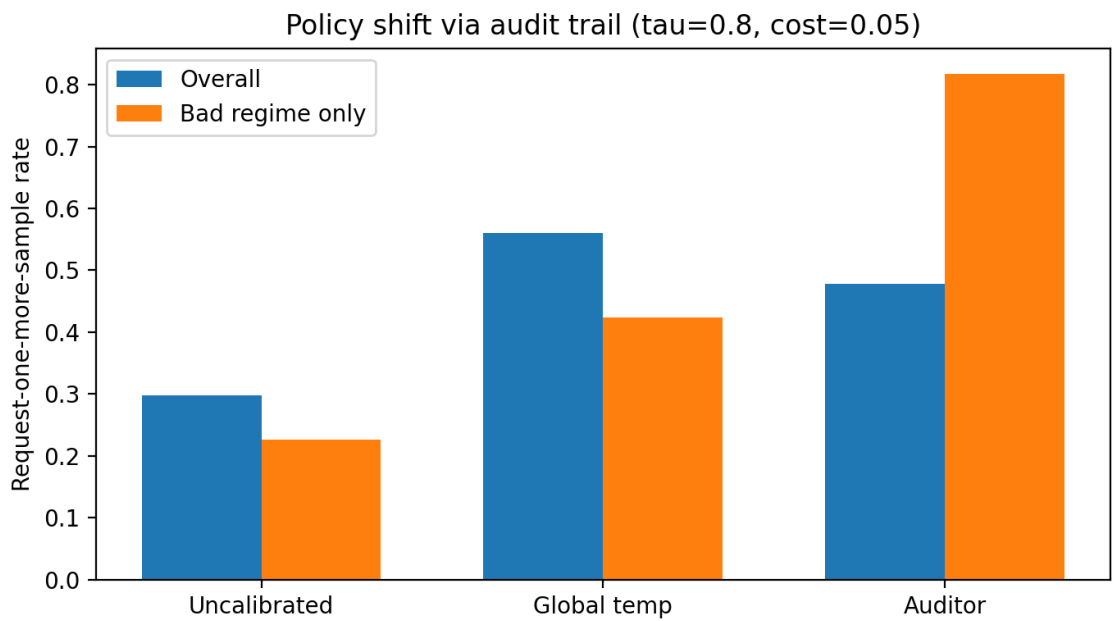

*Figure 3. Control policy effect.* *Request-one-more-sample rates under the act-versus-sample policy ($\tau = 0.8$, $\kappa = 0.05$), shown overall and for the bad-regime trials.*

Figure 3 shows how calibration differences become behavioral under the confidence-threshold policy. The auditor shifts information-seeking toward the degraded regime, requesting additional evidence far more frequently when support is weak. In contrast, content-dominated mappings either under-sample under miscalibration or sample broadly across conditions, reflecting the limits of a single global confidence mapping under regime shift.

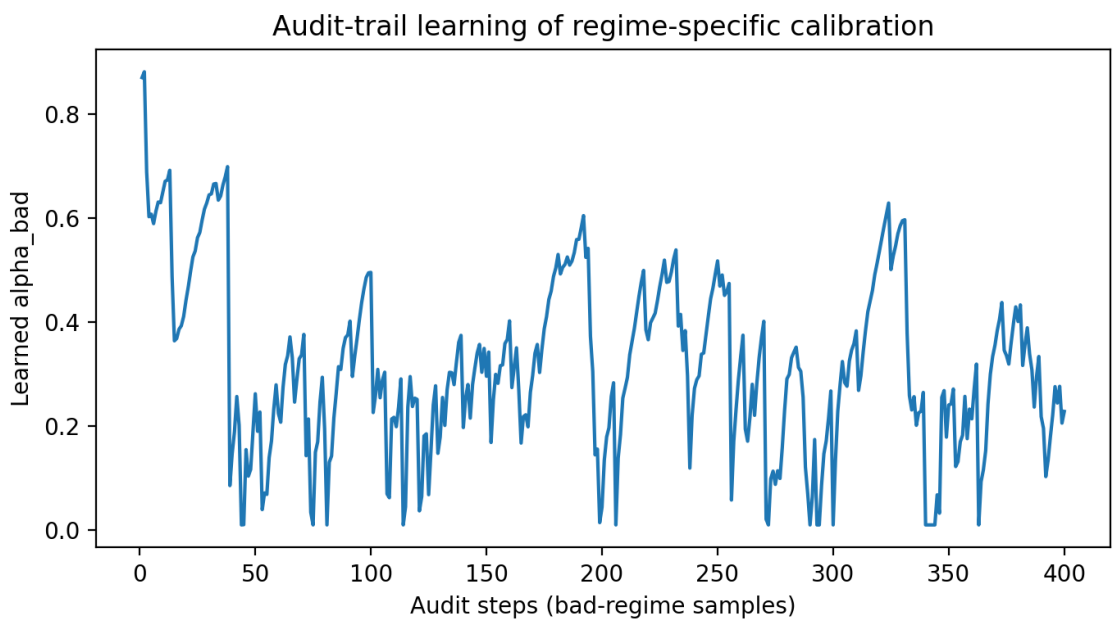

*Figure 4. Audit-trail learning.* Online adaptation of the bad-regime calibration parameter $\alpha_{bad}$ over successive bad-regime outcomes (sample-wise SGD).

Figure 4 illustrates an online variant in which $\alpha_{bad}$ is updated from outcome feedback on bad-regime trials. The trajectory is noisy because updates are stochastic and sample-wise, but it demonstrates that regime-specific calibration need not be hand-set. Repeated prediction-outcome pairs can adapt the confidence mapping in the degraded regime, producing systematic changes in confidence and downstream control.

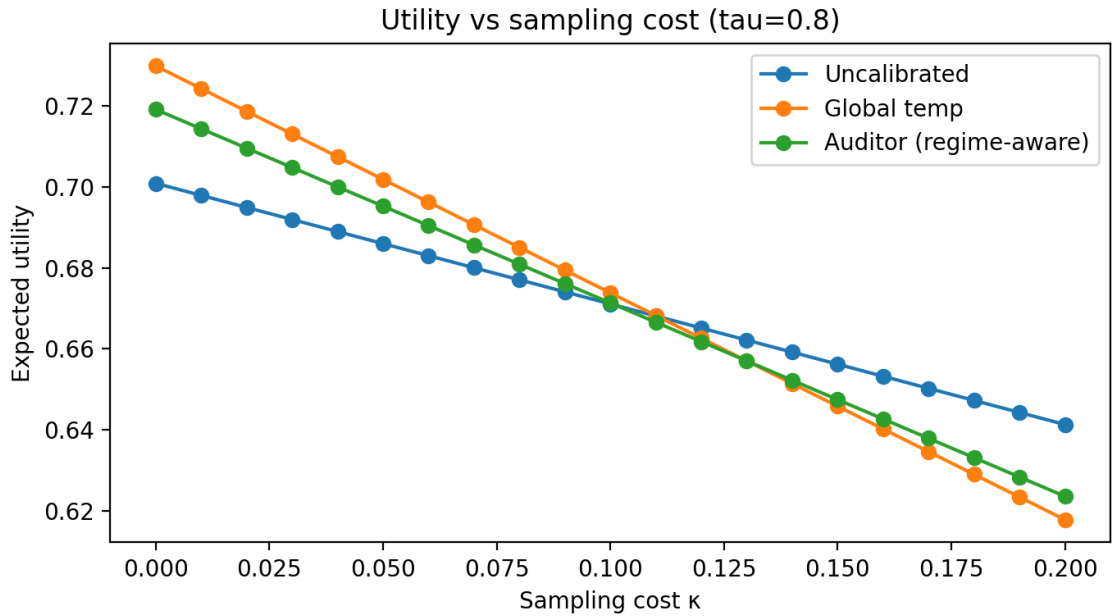

**Figure 5. Utility vs sampling cost.** Expected utility under the act-versus-sample policy as the sampling cost $\kappa$ varies ($\tau = 0.8$).

Figure 5 makes the robustness-cost tradeoff explicit by sweeping the sampling cost $\kappa$. As $\kappa$ increases, models that request additional samples more frequently incur larger penalties, producing different slopes and potential crossovers in expected utility. The auditor's support-conditioned sampling improves calibration and degraded-regime performance, but can reduce utility when sampling is expensive, illustrating a clear tradeoff between robustness under shift and evidence cost

Overall, these results support the core dissociation. When a controller can condition confidence on compact support context, it achieves substantially improved calibration under regime shift and exhibits targeted information-seeking behavior that is not captured by a single global confidence mapping.

**4. Discussion**

This simulation provides a minimal computational realization of the companion framework's claim that preserving support structure for global reuse can induce dissociations between content performance and higher-order behavior. The auditor's advantage does not arise from a different content decision rule. It arises from access to a compact support summary, here instantiated as a binary regime flag, that allows confidence calibration to be conditioned on reliability context and thereby alters downstream control.

Under regime-dependent miscalibration, a content-dominated mapping can continue to convert evidence strength into confidence as if reliability were stationary. The result is that the system can remain confidently wrong in the degraded regime and fail to request additional evidence when it is most valuable. By contrast, a support-aware auditor adapts the confidence mapping to context, which in turn shifts the act-versus-sample policy toward information-seeking under low-support conditions.

Two points are especially relevant for the broader theoretical proposal. First, regime-conditioned calibration is an explicit implementation of an audit trail. Outcome feedback accumulates into a learned mapping from support conditions to appropriate confidence. Second, coupling confidence to policy makes these differences behaviorally measurable. Even with a shared content decision computation, changes in support availability produce systematic differences in information-seeking and cost-robustness tradeoffs. This provides a computational analogue of the claim that manipulations of support preservation can alter control and report-related behavior even when content discrimination is held approximately fixed.

The model is intentionally simplified. Real biological systems would represent support structure with richer, continuous summaries rather than a binary regime indicator, and the mapping from support to calibration would likely be implemented by distributed circuitry rather than a single temperature parameter. Nonetheless, the toy model isolates the key mechanism. Globally reusable support summaries can shape system-level confidence and policy, and audit-trail learning can produce systematic, history-dependent changes in control under reliability shift.

## 5. Limitations and future work

This study uses a deliberately minimal generative process and a stylized calibration mechanism. The support summary is instantiated as a binary regime indicator, standing in for richer, continuous support variables such as coherence, drift, distortion signatures, or provenance. Future work can extend the model to multi-class content, richer support summaries, and active policies that choose among sensors or allocate attention. A particularly relevant extension is to remove direct access to the true regime variable and instead require the auditor to infer support conditions from statistics of the evidence stream, thereby testing whether support summaries can be learned and made globally reusable rather than provided.

## 6. Conclusion

We provided a minimal computational test of the claim that globally reusable support structure can induce dissociations between content inference and system-level confidence and control. In a two-channel cue-integration task with regime shift, an auditor that conditions calibration on a compact support summary improves probabilistic calibration, especially in the degraded regime, and produces distinct information-seeking behavior under an act-versus-sample policy. These results show that support-preservation hypotheses yield measurable behavioral signatures in simple simulations and motivate richer model-based implementations.

## Appendix A. Parameter settings

Trials: $N = 250{,}000$ (50% train / 50% test). Regime probability: $p_{bad} = 0.3$. Noise: $\sigma_A = 0.7$; $\sigma_{B,good} = 0.7$; $\sigma_{B,bad} = 2.0$. Channel B assumed noise for local log-likelihood computation: $\sigma_{B,assumed} = 0.7$. Policy: confidence threshold $\tau = 0.8$; sample cost $\kappa = 0.05$. Calibration fitting: grid search over $\alpha \in [0.05, 5.0]$ with 300 points. Audit-trail learning illustration: SGD on $\alpha_{bad}$ with learning rate 0.05 for 400 bad-regime samples.

**Funding:**

There are no funders to report for this submission.

**Data:**

The data underlying this article will be shared on reasonable request to the corresponding author.